\documentclass[preprint,showpacs,preprintnumbers,amsmath,amssymb]{revtex4}
\usepackage{graphicx}% Include figure files
\usepackage{dcolumn}% Align table columns on decimal point
\usepackage{bm}% bold math

\begin{document}

%\preprint{APS/123-QED}

\title{A Twisted Pair Cryogenic Filter}% Force line breaks with \\

\author{Lafe Spietz}
 \email{lafe.spietz@yale.edu}
\author{John Teufel}
\author{R. J. Schoelkopf}%
\affiliation{Departments of Applied Physics and Physics, Yale
University}

\date{\today}% It is always \today, today,

\begin{abstract}
In low temperature transport measurements, there is frequently a
need to protect a device at cryogenic temperatures from thermal
noise originating in warmer parts of the experiment. There are
also a wide range of experiments, such as high precision transport
measurements on low impedance devices, in which a twisted-pair
wiring configuration is useful to eliminate magnetic pickup.
Furthermore, with the rapid growth in complexity of cryogenic
experiments, as in the field of quantum computing, there is a need
for more filtered lines into a cryostat than are often available
using the bulky low temperature filters in use today. We describe
a low cost filter that provides the needed RF attenuation while
allowing for tens of wires in a twisted pair
configuration with an RF-tight connection to the sample holder.  Our filter consists of manganin twisted pairs wrapped in copper tape with a light-tight connection to the shield of the sample holder.  We demonstrate agreement of our filter with a theoretical model up to the noise floor of our measurement apparatus (90 dB).  We describe operation of our filter in noise thermometry experiments down to 10 mK. 
\end{abstract}

\maketitle

\section{\label{sec:level1}Introduction}

    Nanoelectronic devices such as single electron transistors, quantum dots, and other quantum circuits
are sensitive to thermal radiation from the portions of the
experiment which are above the base temperature of the cryostat
\cite{devoretJAP}. Any thermal radiation that is not filtered can
heat the electrons in the device being measured, defeating the
purpose of cryogenic transport measurements. Furthermore, in some
experiments, such as switching statistics measurements of
Josephson junctions \cite{Martinisdevoretclarke}, very small
numbers of microwave photons can ruin the experiment. Thus, there
is a need in cryogenic electronic experiments for filters that
both provide thermalization to very low temperatures and have very
high attenuation throughout a wide range of frequencies.

    Filtering for single electron circuits is typically accomplished
by use of a powder filter at the base temperature of the cryostat
\cite{Martinisdevoretclarke}, which consists of a copper tube
packed with metal powder (either stainless steel or copper) into
which approximately a meter of wire is coiled.  Each end is
typically fitted with a coaxial connector (e.g. SMA,) and the
filters are connected using semi-rigid coaxial cables.  A major
drawback of this technique is that the space taken up per wire and
the quantity of copper that must be cooled to base temperature
(typically about 90 grams/wire) becomes prohibitive for large
numbers of wires.  Because these filters rely on SMA connectors
and semi-rigid cables, there a limited number that can be
practically installed in a given cryostat, and they must always be
installed in a single-ended configuration. Under certain
circumstances, it can be very advantageous to measure devices in a
twisted pair configuration \cite{ott}, and the single-ended SMA
filters are not as compatible with that as a true twisted pair
filter.

    In addition to the standard powder filters, there exist a variety of
other cryogenic filters
\cite{devoretJAP,courtois_RSI,zorin_RSI,kevin,fukushima}. While these are
not all as intrinsically bulky as the powder filters, they all
share the drawbacks of being built with coaxial connectors in a
single-ended configuration.  Furthermore, the
photolithographically defined filters
\cite{devoretJAP,courtois_RSI} add significant cost and complexity
to the design.

    We have constructed and tested a filter that can be easily
hand-made from off-the-shelf components. Because the filter is in
the form of a flexible cable, it can be bent to any desired shape
and can easily fit in a cryostat wherever space is available.
Unlike the filters that use SMA cables, our filter can be fitted
with any one of a number of many-pin connectors and requires very
little effort and an insignificant cost in space to increase the
number of wires used in the design.

\section{Description of Filter}

Our filter consists of a twisted pair of resistive wires wrapped
tightly in copper tape.  The copper tape acts both as a shield and
as a ground to which the lines are capacitively coupled.  The
copper tape is pressed around the wires very closely so that the
distance between the wires and the shield is that of the insulator
thickness on the wires, about 10 microns.  Thus there is a high
capacitance from the wires to ground as well as between the two
wires in any given pair. This capacitance to ground combined with
the high resistivity of the wires makes the cable a continuous RC
line (see figure 2).  This system is simple enough that the RF
loss of the line as a function of length can be fit to a simple model.  This simplicity could prove useful when trying to model the time domain behavior of fast voltage pulses down the line.

The ends of the wires are soldered to a short segment of ribbon
cable which allows for easy application of connectors and which is more
mechanically and thermally robust than the main section of the
cable (see figure 1.)  The ribbon cable end may then be slotted
through a copper shield into the box which contains the
experiment, and sealed in with low temperature solder.  This
creates RF-tight continuous shielding from the noisy end of the
filter down into the low noise shielded enclosure containing the
experiment.

\begin{figure}
\includegraphics{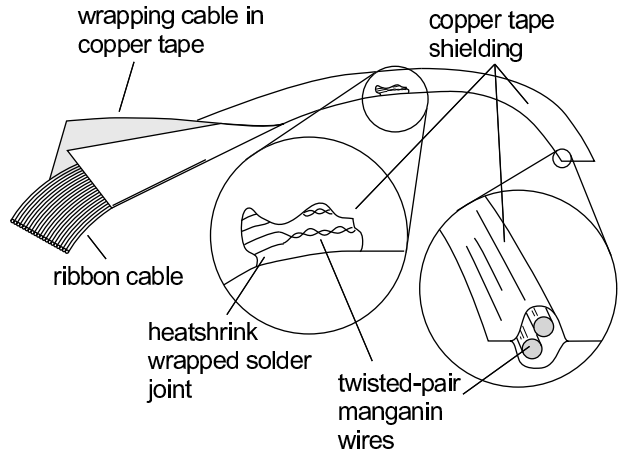}
\caption{\label{fig:assemblydiag}Assembly Diagram.  Blown up views show the transition from
ribbon cable to twisted pair and a cross section of the twisted
pair cable (not to scale.)}
\end{figure}

\section{Theory}
\begin{figure}
\includegraphics{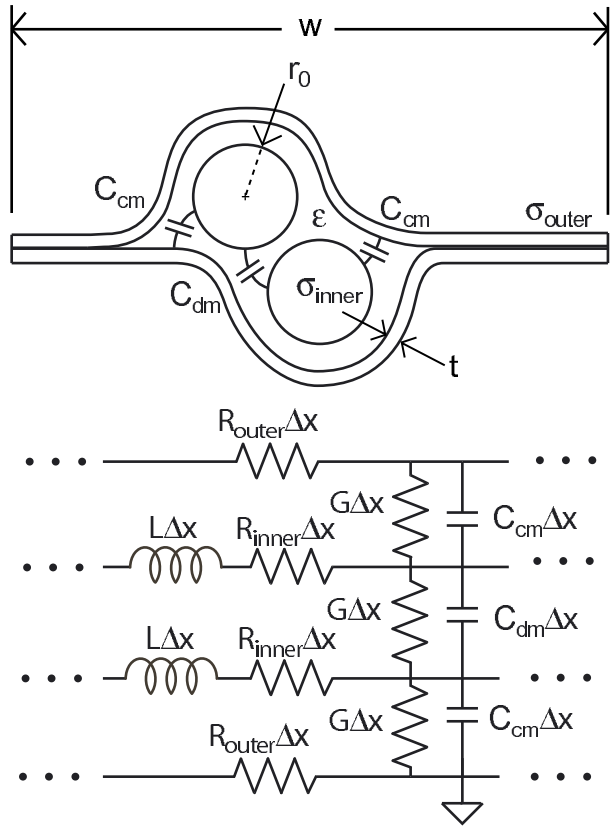}
\caption{\label{fig:RClinediag}Above cartoon shows how the circuit parameters correspond
to the physical components of the filter. The schematic below the
cartoon shows the circuit model used for the theory.}
\end{figure}

We can analyze the filter as a set of transmission lines with
capacitance per unit length $C_{cm}$ and $C_{dm}$ for common mode
and differential mode respectively, resistance per unit length R,
and inductance per unit length L.  Because of the finite skin
depth of high frequency signals, R is frequency dependent.  To
analyze this line, we follow the standard treatment of the lossy
transmission line \cite{pozar} to find the complex propagation
constant $ \gamma = \alpha + i\beta = \sqrt{(R(\omega)+i\omega
L)(G(\omega) + i\omega C)}, $ which corresponds to travelling wave
solutions $V(x)=V_0^+e^{-\gamma x} + V_0^-e^{\gamma x}. $

    $R(\omega)$ can be computed from the DC conductivity
$\sigma$ and the fraction of the cross-sectional area that carries
current at frequency $\omega$. The skin depth is $
\delta=\sqrt{2/\mu_0\sigma\omega}, $ where $\mu_0$ is the
permeability of the vacuum and $\sigma$ is the DC conductivity of
the material. We then compute the resistance of the inner and
outer conductors at finite frequency as follows: $
R_{inner}=1/2\pi\sigma_{inner}\delta_{inner}^2(e^{-r/\delta_{inner}}+(\frac{r_0}{\delta_{inner}})-1)
$, and $
R_{outer}=1/2\pi\sigma_{outer}\delta_{outer}(1-e^{-t/\delta_{inner}})
$.

To find the capacitances per unit length we both compute
theoretical values and measure capacitances directly.  The
measurement of the capacitance is described below in the section
on filter testing.  The theoretical models are based on various
published descriptions of transmission lines \cite{wadell}.

Both the common mode and differential mode are treated using the
simplified model shown in figure 3, with a single value for C
calculated from the series and parallel combinations of $C_{dm}$
and $C_{cm}$.  We can compute C as follows:

\begin{equation}
 C=\left\{ \begin{array}{l@{\quad \quad}l}
\frac{C_{dm}C_{cm}}{C_{dm}+C_{cm}}+C_{cm}&\mathrm{for\ common\ mode} \\
C_{dm}+\frac{C_{cm}}{2}& \mathrm{for\ differential\ mode}
\end{array}
 \right.
.\end{equation}

The total resistance R used in the final calculation is then found
from
\begin{equation}
 R=\left\{ \begin{array}{l@{\quad \quad}l}
R_{inner}+R_{outer}&\mathrm{for\ common\ mode} \\ 2R_{inner}&
\mathrm{for\ differential\ mode}
\end{array}
 \right.
.\end{equation}

The shunt conductance as a function of frequency is computed from
the loss tangent $\tan\delta$ as follows:
\begin{equation}
G(\omega)=\omega C \tan\delta,
\end{equation}
where $\tan\delta$ is the loss tangent of the dielectric material.

  The capacitance per unit length $C$ and the dielectric constant
$\epsilon$ are fit parameters in the model, and the inductance $L$
is calculated from them using the assumption that
$L=\epsilon\mu_0/C$.  $\epsilon$ may be estimated from the value
for the polyimide insulation on the wire.

\begin{figure}
\includegraphics{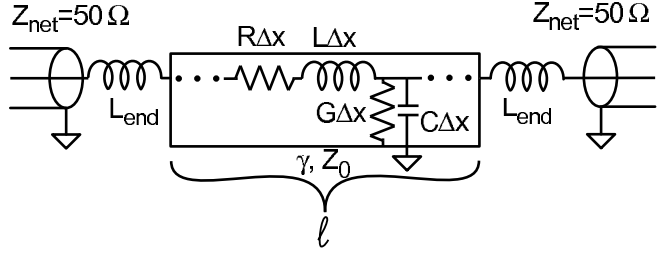}
\caption{\label{fig:txlinecartoon}Cartoon of lossy transmission line with inductances at
each end.}
\end{figure}

The lumped-element inductance at each end of the line is from the
wire that connects the inner conductor of the filter with the
inner conductor of the 50 $\Omega$ line from the network analyzer.
This can be roughly estimated by assuming that the wire has about
a nH per millimeter of inductance, and measuring the length of
bare wire.  Since there are about two or three centimeters of bare
wire at each end, this gives a rough estimate of $L_{end}$ of
about 30 nH.  Not surprisingly, the best-fit inductance turns out
to be a little higher, 45 nH.

With the relevant parameters for the transmission line known, we
can proceed to calculate the propagation constant as described
above.  From this we may use standard transmission line theory to
calculate the insertion loss of the filter cable \cite{pozar}:
\begin{equation}
\mathrm{insertion \
loss}=|S_{12}|^2=|S_{21}|^2=\left|\frac{1}{\cosh(\gamma
l)+\frac{1}{2}(\frac{Z_0}{Z_{net}} +
\frac{Z_{net}}{Z_0})\sinh(\gamma l)}\right|^2,
\end{equation}
where $Z_{net}$ is the impedance of the line from the network
analyzer in the test circuit (50 $\Omega$), and $Z_0$ is the impedance of the transmission line.  For simplicity, we have shown the insertion loss of the filter cable alone.  The
actual theory we compare to network analyzer data includes the end
inductances.

\section{Assembly and Installation}

The first step in fabrication of the cable is assembling the
twisted pairs by fastening the pair at one end and twisting the
opposite end with an electric drill motor until the pair is
twisted at the desired pitch \cite{webtutorial}. The inner
conductor of our filter is polyimide-coated manganin wire 4 mils (100 microns)
in diameter with a 0.25 mil (6 microns) thick insulating layer
\cite{manganinwire}. Once enough twisted pairs of the right length
have been made, they must be arranged in the order in which they
will be assembled into the cable and soldered to the ends of the
ribbon cable.  The solder joints are individually encased in heat
shrink tubing to prevent shorting.  The entire assembly of wires
and ribbon cables is then wrapped up in copper tape
\cite{coppertape} by folding the tape twice as shown in figure 1.
Once the tape is stuck to itself, form fitting adhesion to the
wires may be achieved by pressing the whole assembly with a
rolling pin.  This gives the cable the profile shown in the
cutaway blowup diagram in figure 1, maximizing the capacitance
between the inner conductor and the shield.

\begin{figure}
\includegraphics{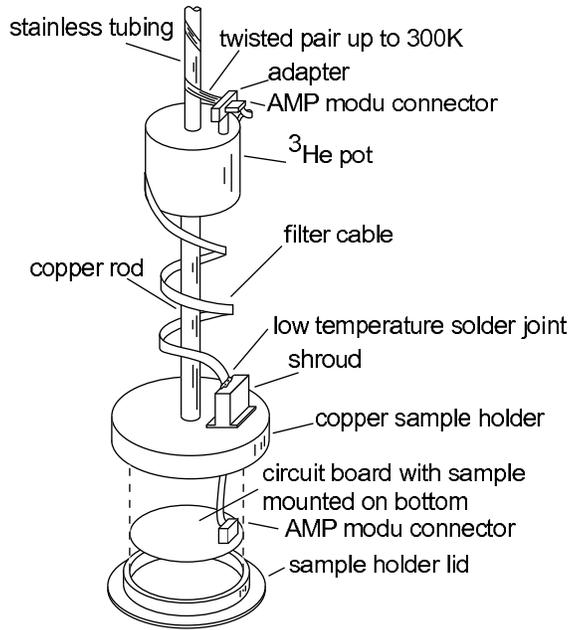}
\caption{\label{fig:installationdiag}Sketch of cable filter installation in a cryostat.  The
cable may be wrapped around the cryostat and bent to shape to fit
into the available space.}
\end{figure}

    The exposed ends of the cable are both ribbon cable, and are
easily fitted with AMP modu solderless connectors.   AMP sells a
wide variety of connectors in the modu line including conversions
to various other kinds of connectors and PC board pluggable
connectors.  This allows for easy connection to whatever connector
one may have in the cryostat as well as to the PC board on which
the sample is mounted.

    In order to make a light-tight connection from the shield of the
cable to the sample holder, we use a copper ``shroud'' through
which the cable is threaded which mates to the sample holder. This
``shroud'' consists of a copper box with a thin slot at one end
and an open face surrounded by a flange at the other end (see
figure 4.)  The copper-shielded ribbon cable part of the cable is
passed through the slot in the shroud which is then sealed with
Indium-Cadmium low temperature solder (applied with a heat gun).
This is done in such a way as to leave a little extra length in
the cable so the AMP modu connector can be mated with the circuit
board inside the sample holder.  The shroud is then bolted over an
opening in the sample holder and sealed with an Indium o-ring.
When the whole system is assembled, there is light-tight shielding
from the beginning of the cable all the way down to the sample
holder, surrounding the sample.

\section{Testing of Filter}

    To test the filter, a test apparatus was constructed that consisted of a
pair of SMA connectors whose center pins were soldered to a pair
of wires.  The SMA connectors were inset in a groove over which a
brass plate was mounted, sealed with a pair of indium o-rings
sandwiching the cable.  This provided a light-tight seal from one
pair of SMA connectors to the other.  The transmission as a
function of frequency was then measured on a HP8753D network
analyzer from 30 kHz to 6 GHz, and from 50 MHz to 40 GHz on a
HP8722D network analyzer. Figure 5 shows a plot of $S_{12}$ at
room temperature compared with theoretical transmission for two
different lengths of cable.  Note that the behavior of the filters
is monotonic and very close to the predicted behavior.

\begin{figure}
\includegraphics{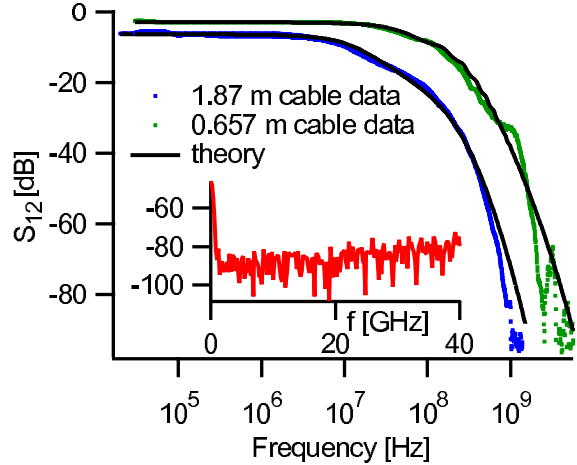}
\caption{\label{fig:Sparam}Throughput vs. frequency for different lengths of cable,
with fit. Note that we only measure directly the common mode
transmission and must rely on the model and the measured
capacitance for the differential mode.  The inset shows the range
up to 40 GHz, which is all at the noise floor above the scale
of the 6 GHz data.}
\end{figure}

    By adjusting the parameters slightly, good fits can be found
 with very reasonable values for all the parameters.  The values
 of the parameters are in Table 1.

\begin{table}\centering
 \caption{\label{table 1}Table of values describing the filter.  L, and C are the inductance and capacitance per unit length of the transmission line, respectively.  $\tan(\delta)$ and $\epsilon_r$ are the loss tangent and relative dielectric constant of the dielectric, respectively.  w and t are the total width of the cable and the thickness of the copper tape, respectively.  $L_{end}$ is the inductance at the end of the test apparatus.}
\begin{tabular}{|c|c|c|}\hline
Parameter & Value & Units\\
\hline\hline $\sigma_{\textrm{manganin}}$&$2.2\times10^6\
$&$\Omega^{-1}m^{-1}$\\\hline
$\sigma_{\textrm{copper}}$&$5.9\times10^7\
$&$\Omega^{-1}m^{-1}$\\\hline $C$&$210\ $ &$\textrm{pF/m}$\\\hline
$L$&$190\ $&$\textrm{nH/m}$\\\hline $\tan(\delta)$&$0.01$&\\\hline
$\epsilon_{r}$&$3.5$&\\\hline $w$&$0.02\ $&$\textrm{m}$\\\hline
$t$&$1\times10^{-4}\ $&$\textrm{m}$\\\hline $L_{end}$&$45\
$&$\textrm{nH}$\\ \hline
\end{tabular}
\end{table}

    Note that $\epsilon_{r}$ is approximately the value of
the relative dielectric constant for polyimide.  In this model,
all of the parameters we use are physically reasonable, which may
be contrasted to the powder filter, in which it is difficult to
construct such a model.

    The capacitance per unit length was measured by connecting a
10 k$\Omega$ resistor in parallel with the cable to a EG$\&$G 5113
low noise pre-amp in front of a Stanford Research Systems SR760
FFT analyzer.  Looking at the RC rolloff of the noise spectrum, it
was possible to determine the capacitance per unit length. By
connecting the 10 k$\Omega$ resistor either between the two
conductors of a twisted pair or from one conductor to the shield
it was possible to separately measure the common mode and
differential mode capacitance.

    By neglecting the resistance of the manganin wire and treating
the whole cable as just a capacitor, the Johnson noise spectrum
can be measured and fit to extract the desired capacitances.  In
order to eliminate the extra capacitance from the connectors at
the end of the cable, we measure different cable lengths and
estimate the capacitance per unit length by computing $\Delta
C/\Delta l$.  This yielded capacitance values of C$_{cm}$=200 pF/m
and C$_{dm}$=51 pF/m.

    The capacitances can also be estimated to check how reasonable
our values are by computing certain theoretical limits.  In one
limit, we consider the line to be a coaxial line with the
insulation on the wire surrounded tightly by copper. In the
opposite limit, we consider the wire to be sandwiched halfway
between a pair of ground planes pushed up against the outside of
the insulation.  In the first limit, the capacitance is given by
\begin{equation}
C=\frac{2\pi\epsilon_0\epsilon_{r}}{\ln(\frac{r+s}{r})}=1.6\
\textrm{nF/m},
\end{equation}
where $\epsilon_{r}$ is the relative dielectric constant, r is the
radius of the wire (50 microns), and s is the thickness of the
insulation (6.4 microns.)  In the second limit, the capacitance of
the so-called ``slabline'' \cite{wadell} can be calculated as
\begin{equation}
C=\frac{2\pi\epsilon_0}{\ln(\frac{4(r+s)}{\pi r})}=150\
\textrm{pF/m},
\end{equation}
where we assume that the average relative dielectric constant is
1.  As expected, the actual capacitance, 210 pF/m, is between
these limits.

    Having installed the cable in the cryostat, we have tested it by
measuring superconducting tunnel junctions and single electron
transistors.  By observing the single electron transistor
current-voltage characteristics, we can evaluate the cable's
filtering capability.  So far we have used this cable in our
Heliox pumped $^3$He cryostat and our dilution refrigerator with
great success for both tunnel junction experiments and single
electron transistor experiments. In addition, we test the
thermalization of the electrons at low temperatures by measuring
the noise of a tunnel junction as a function of bias voltage, i.e.
by doing a shot noise thermometer measurement
\cite{spietzscience}.  We used this method to verify that the
electron temperature of our sample was as low as 10 mK \cite{SNT_lowtemp}.  In all samples tested, the device physics observed was consistent with an electron temperature equal to the physical temperature of the cryostat.

\newpage

\begin{acknowledgments}
We thank Michel Devoret for useful discussions, and the David and
Lucile Packard Foundation for support.
\end{acknowledgments}
\newpage


\newpage

\end{document}